\documentclass[11pt,a4paper]{article}
\usepackage[T1]{fontenc}
\usepackage[utf8]{inputenc}
\usepackage{authblk}
\usepackage{booktabs} 
\usepackage[english]{babel}
\usepackage[compress]{cite}
\usepackage[letterpaper,top=2cm,bottom=2cm,left=3cm,right=3cm,marginparwidth=1.75cm]{geometry}

\usepackage{amsmath}
\usepackage{graphicx}
\usepackage{color}
\usepackage{lineno}
\usepackage[colorlinks=true, allcolors=blue]{hyperref}

\title{Lean classical-quantum hybrid neural network model for image classification}

\author[a]{
Ao Liu,$^{1}$
Cuihong Wen,$^{1}$\thanks{E-mail: cuihongwen@hunnu.edu.cn}
Jieci Wang,$^{2}$\thanks{E-mail: jieciwang@hunnu.edu.cn}

$^{1}$College of information science and engineering, Hunan Normal University, Changsha, 410081, China\\
$^{2}$Department of Physics and Key Laboratory of Low Dimensional Quantum Structures and Quantum Control of Ministry of Education, Hunan Normal University, Changsha, 410081, China
}
\date{} 

\begin{document}
\maketitle

\begin{abstract}
The integration of algorithms from quantum information with neural networks has enabled unprecedented advancements in various domains. Nonetheless, the application of quantum machine learning algorithms for image classification predominantly relies on traditional architectures such as variational quantum circuits. The performance of these models is closely tied to the scale of their parameters, with the substantial demand for parameters potentially leading to limitations in computational resources and a significant increase in computation time. In this paper, we introduce a Lean Classical-Quantum Hybrid Neural Network (LCQHNN), which achieves efficient classification performance with only four layers of variational circuits, thereby substantially reducing computational costs. Our experiments demonstrate that LCQHNN achieves 100\%, 99.02\%, and 85.55\% classification accuracy on MNIST, FashionMNIST, and CIFAR-10 datasets. Under the same parameter conditions, the convergence speed of this method is also faster than that of traditional models. Furthermore, through visualization studies, it is found that the model effectively captures key data features during training and establishes a clear association between these features and their corresponding categories. This study confirms that the employment of quantum algorithms enhances the model's ability to handle complex classification problems.
\end{abstract}

\section{Introduction}

Neural Networks (NNs), as a vital machine learning tool, have achieved widespread success across various domains \cite{Jordan2015MachineLearning, Sun2023FinetuningGN,kingma2013auto,goodfellow2014generative, ho2020denoising,Rumelhart1986LearningRepresentations}. Extensive research has demonstrated the formidable capabilities of neural networks in pattern recognition, feature extraction, and classification decision-making \cite{Dosovitskiy2020AnII, he2015deepresiduallearningimage,Hochreiter1997LongShortTermMemory,Subakan2021Attention}. However, the performance of traditional deep learning algorithms is intimately linked to the scale of the model and the size of the dataset. An increase in the scale of parameters can lead to constraints in computational resources and a significant escalation in computation time. Moreover, the optimization of hyperparameters presents a challenge, as there is currently no universally effective optimization method akin to gradient descent for hyperparameter optimization. When the parameter scale is large, the temporal cost of evaluating a set of hyperparameter configurations becomes exceedingly high.
Recently, there has been a notable increase in interest regarding the integration of neural networks with quantum technologies \cite{Dunjko:2016naw, Tamiya2022, Jager2023, Fan2024, Wang2024, p2023, Tian2022}. Quantum algorithms possess a range of distinctive advantages that have the potential to significantly enhance both the performance and efficiency of traditional computational methods\cite{Gong_2024,9647979, 10254235}. For instance, Liu et al. proposed a hybrid quantum-classical convolutional neural network (QCCNN), which employs parametric quantum circuits to replace classical convolutional kernels, enabling the model to better extract patterns from complex data distributions\cite{Liu2021}. Similarly, A. Senokosov et al. introduced parallel quantum layers (HQNN-Parallel) and quantum convolutional layers (HQNN-Quanv), demonstrating that hybrid models can outperform classical counterparts even with fewer trainable parameters\cite{Senokosov_2024}. However, the same study also highlights that having fewer trainable parameters does not inherently guarantee more efficient execution in quantum machine learning (QML) models compared to classical ones. This limitation arises from the significantly higher computational overhead associated with quantum model training and inference.

Variational Quantum Circuits (VQCs) constitute a class of quantum algorithms designed to solve optimization and machine learning problems through tunable parameterized quantum circuits \cite{Biamonte2017QuantumMachineLearning,Qi2023,Bagoun2024,Bhowmik2024}. Their core mechanism relies on optimizing a limited set of quantum gate parameters (typically maintaining controllable parameter scales) rather than relying on large-scale parameter accumulation to approximate optimal solutions. VQCs extract feature patterns through quantum entanglement hierarchies that are undetectable by classical neural networks \cite{Du2021PRXQuantum}, thereby significantly enhancing robustness against classical adversarial attacks and providing theoretical foundations for practical quantum advantages in machine learning \cite{PhysRevResearch.5.023186}. Recent advancements in quantum machine learning have witnessed breakthroughs in two key technological pathways. On one hand, in the realm of hybrid quantum-classical feature engineering, Chen et al. \cite{chen2023} innovatively combined the multi-scale entanglement renormalization Ansatz with box-counting fractal feature extraction to construct a Quantum Convolutional Neural Network (QCNN). This network has demonstrated superior accuracy compared to classical Convolutional Neural Networks (CNNs) in medical imaging classification tasks for breast cancer. Moreover, the universality of this quantum feature engineering approach has been further validated in the field of high-energy physics. In particle collision event classification tasks \cite{Chen2022QuantumConvolutionalNeuralNetworks}, Chen et al. found that, with a comparable number of parameters, QCNNs outperformed classical CNNs \cite{Krizhevsky2012ImageNetClassification} in both test accuracy and learning speed. On the other hand, to address the resource constraints of Noisy Intermediate-Scale Quantum (NISQ) devices, Zhou et al. proposed a hybrid quantum-classical Generative Adversarial Network (QGAN) \cite{Zhou2022HybridQG}. This network innovatively employs image remapping techniques to transform complex multi-modal distributions into single-modal representations that are processable by quantum systems. The scheme successfully achieved high-quality reconstruction of MNIST and Fashion-MNIST images, marking a significant step toward the real-world deployment of quantum technologies.

Although the participation of quantum algorithms can help models efficiently complete image classification tasks with limited computing resources, there are still many difficulties and limitations in specific applications \cite{McClean2018BarrenPlateaus,chakrabarti2019quantum}, mainly from the following two aspects. The 'curse of dimensionality' poses a significant challenge in implementing quantum machine learning algorithms for image classification tasks \cite{PhysRevLett.122.040504, doi:10.1126/science.abn7293, Zoufal2019}. Image data typically exhibits extremely high spatial dimensions, and encoding such data into quantum states requires substantial quantum bit resources. Specifically, a system composed of n qubits has a state space dimension of $2^{n}$, which implies that the quantum resources required for VQCs grow exponentially with increasing data dimensionality, far exceeding the processing capabilities of current NISQ hardware. Meanwhile, complex quantum systems are highly sensitive to noise, and increasing the depth of VQCs can lead to more noise accumulation. For instance, Wang et al. (2021)\cite{wang2021} demonstrated that noise in NISQ devices induces barren plateaus in the training landscape (i.e., vanishing gradients). As the circuit depth of VQAs scales linearly with the number of qubits $n$, the parameter gradients vanish exponentially with $n$. In a parallel study, Aidan Pellow-Jarman et al. investigated the optimal circuit depth for the quantum approximate optimization algorithm (QAOA) and revealed that each additional layer introduces hardware noise \cite{Aidan2024}. Their experiments showed that task performance peaks at approximately six layers. Beyond this critical depth, noise accumulation dominates, leading to a gradual performance degradation. This noise-depth trade-off can be extended to high-dimensional tasks such as image classification. These studies indicate that the noise characteristics of current NISQ devices severely limit the scalability of quantum algorithms. Therefore, how to enhance algorithmic efficiency while ensuring the stability and reliability of quantum computing in image classification remains a critical challenge to be addressed.

The development of effective quantum image classification models and the minimization of noise impact are critical research areas in machine learning. To realize quantum enhancement of image classification with limited hardware resources, this paper introduces a lean classical-quantum hybrid neural network (LCQHNN) for image classification. We employ classical multi-channel convolutional operators for nonlinear feature extraction from images. Compared to quantum convolution kernels, this approach offers superior hardware compatibility, ensuring better alignment with contemporary computing architectures. Meanwhile, we introduce a streamlined four-layer variational quantum circuit (VQC) module that effectively reduces quantum circuit complexity and mitigates noise sensitivity. By minimizing circuit depth and optimizing parameter sharing, our approach enhances stability and reliability, avoiding the decoherence errors and resource overhead often associated with parallel quantum circuits. Experimental results on public datasets demonstrate that LCQHNN significantly outperforms convolutional neural networks (CNNs) in terms of convergence speed and classification accuracy.
Specifically, it achieved training convergence speeds that were 75\% and 70.59\% faster than those of CNNs with the same parameters on the MNIST and FashionMNIST datasets, while also achieving excellent accuracy on the test sets. In addition, we utilize Grad-CAM heat maps to visualize the prediction process of LCQHNN, revealing how it learns edge contours without prior knowledge. These visualizations enhance our understanding of the model's decision-making process and validate its robustness in image recognition tasks.

\section{Methodology}
\subsection{Introduction to the LCQHNN Algorithm}

\begin{figure}
\centering
\includegraphics[width=1\linewidth]{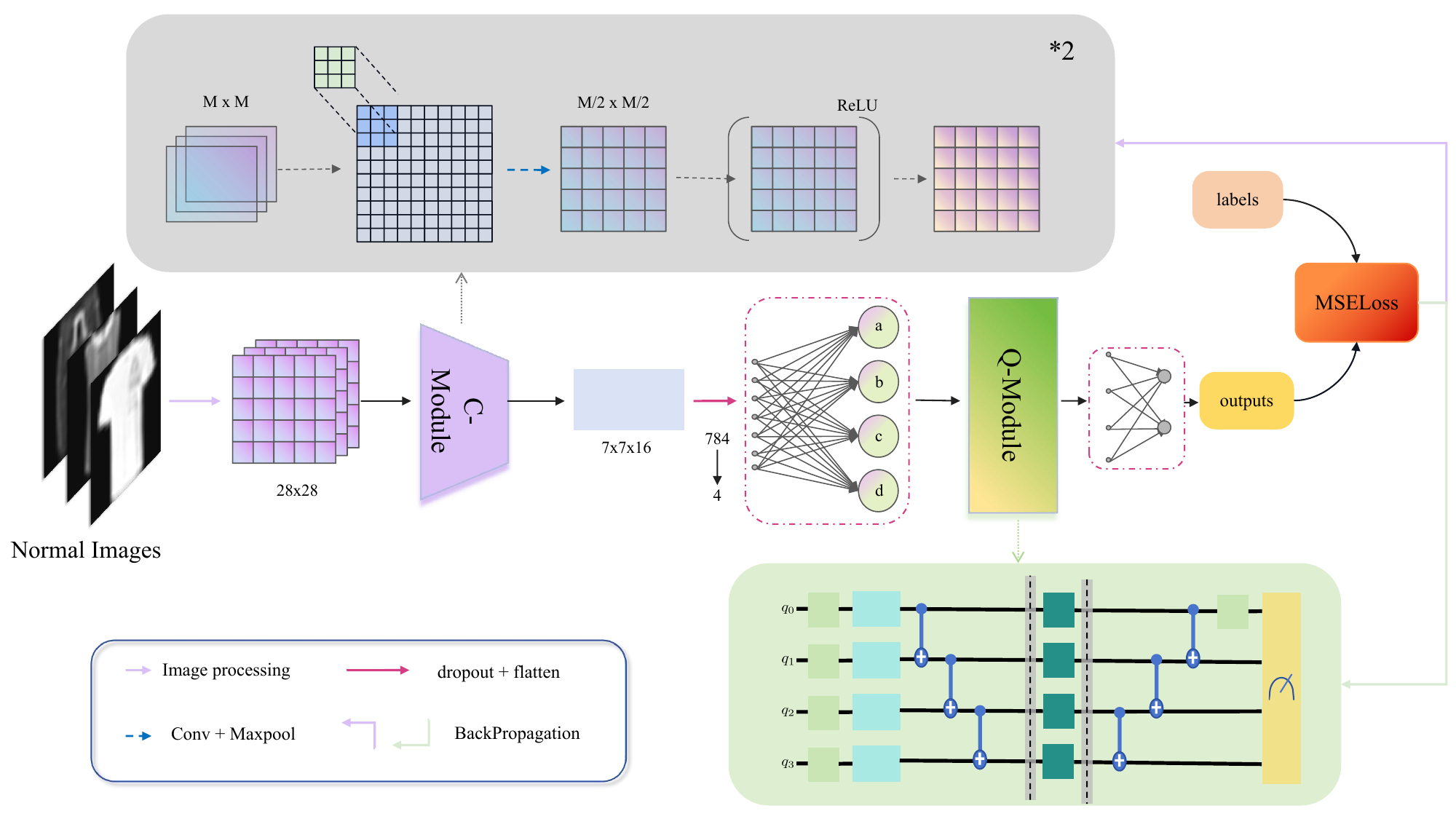}
\caption{\label{fig:1}\textbf{Lean classical-quantum hybrid neural network structure for studying image classification.} Firstly, feature extraction is performed using classical neural networks, and then the extracted features are passed on to VQCs for classification tasks. Finally, the quantum data is converted back to classical data through the measurement layer to obtain the final classification result. }
\end{figure}

Let's first organize our theoretical framework. The quantum machine learning algorithm proposed in this study is a classical quantum classical process. We position the classical convolutional neural network at the front end of VQCs for feature extraction while reducing the spatial dimension to meet the quantum bit requirements of LCQHNN\cite{Araz2022}. Although we have reduced the spatial dimension of the image, CNN maps the information that needs to be focused on to a higher feature dimension by increasing the number of channels, which is used to extract hidden features and improve accuracy\cite{kim2014}. Therefore, the features extracted by the CNN algorithm can be used as inputs for VQCs, where classical state data becomes quantum states. VQCs is the core of our algorithm, similar to the general approximation theorem in Artificial Neural Network. VQCs represents a circuit that can fit the objective function. During the training process, VQCs will repeat the feedforward process and compare the results with the true labels of the training data. By iteratively adjusting the parameters of the variational quantum circuit, the cost function is minimized in terms of the difference between the predicted label and the true label. Finally, the measurement results of the four qubits yield a set of 4-bit classical states. These classical data are then processed through a fully connected neural network for linear transformation, and subsequently normalized via the softmax activation function to produce a probabilistic distribution over all target classes.

Prior to feature extraction, the input images undergo standardized preprocessing to ensure compatibility with the LCQHNN architecture. For the dataset used in the experiment mentioned in this paper, raw images are uniformly processed as follows.
\begin{itemize}
     \item \textbf{Grayscale Normalization}:
     \begin{itemize}
         \item For grayscale datasets (MNIST/FashionMNIST), pixel values are linearly scaled from $[0, 255]$ to $[0.0, 1.0]$:
         \[
         I_{\text{norm}} = I_{\text{raw}} / 255.0
         \]
         \item For RGB datasets (CIFAR-10), each channel is independently normalized:
         \[
         I^{(c)}_{\text{norm}} = I^{(c)}_{\text{raw}} / 255.0,\quad c \in \{R,G,B\}
         \]
     \end{itemize}

     \item \textbf{Spatial Standardization}:
     \begin{itemize}
         \item MNIST/FashionMNIST images are resized to $28 \times 28$ pixels
         \item CIFAR-10 images are resized to $32 \times 32$ pixels
     \end{itemize}

     \item \textbf{Tensor Conversion}:
     \begin{itemize}
         \item MNIST/FashionMNIST: $[1, 28, 28]$ (single channel)
         \item CIFAR-10: $[3, 32, 32]$ (RGB channels)
     \end{itemize}
 \end{itemize}

As shown in Figure \ref{fig:1}, we construct the quantum algorithm model from five aspects to achieve image classification. Firstly, the feature extraction model of LCQHNN consists of two convolutional layers and two pooling layers. The convolutional layers are used to extract information to ensure reliable information transmission throughout the entire network architecture. Reducing the spatial dimension of the pooling layer can decrease the number of parameters and computational complexity in subsequent layers, making the network more efficient. In the second aspect, to ensure non-linear features, we added an activation layer composed of ReLU function between each convolutional layer and pooling layer. This work achieves nonlinear mapping of feature vectors. We assume that $f$ is the input image, $g$ is the convolution kernel, $(i, j)$ are the coordinates of the input image, and $(m, n)$ are the coordinates of the summation. So the convolution result should be $(f * g)(i, j) = \sum_{m=-\frac{k-1}{2}}^{\frac{k-1}{2}} \sum_{n=-\frac{k-1}{2}}^{\frac{k-1}{2}} f(m, n) g(i-m, j-n)$. Where k is the size of the convolution kernel, we set it to 3. The formula for ReLU is $max(0,x)$. We assume that $Window(i, j)$ represents a pooling window centered around $(i, j)$, and $P(i, j)$ is the value of the pooled output feature map at position $(i, j)$. So the extracted feature results \cite{PhysRevC.105.034611} are as follows.

\begin{equation}
     P(i, j) = \max_{(m, n) \in \text{Window}(i, j)} \max(0,\sum_{m=-\frac{k-1}{2}}^{\frac{k-1}{2}} \sum_{n=-\frac{k-1}{2}}^{\frac{k-1}{2}} f(m, n) g(i-m, j-n))
\end{equation}

Thirdly, after feature extraction, we set up a transition model composed of fully connected networks, which reduces the dimensionality of features from 256/400 to 4 to accommodate quantum computing constraints, the fully connected input dimension depends on the size of the initial input image. Due to the property of parameter sharing in convolutional neural networks, neurons in the network may develop complex co adaptation relationships. Therefore, we incorporate Dropout as a regularization technique into the module, which reduces co adaptation relationships among neurons by randomly discarding them. Intuitively described, it can be seen that each Dropout generates a different "sparse" network, which is equivalent to training many different networks and averaging their prediction results, which usually helps prevent overfitting of the training data by the network.

Fourthly, we use quantum encoding technology to transform the 4-dimensional feature vector into a 4-dimensional 4-bit quantum state, followed by a series of carefully designed single bit operations. Finally, the quantum measurement results are processed through a fully connected network and softmax function to obtain the classification outcome. We will provide a detailed description of the quantum algorithm in Section \ref{section:0}.

\subsection{\label{section:0}Variational Quantum Circuit Architecture}

\begin{figure}[ht!]
     \centering
     \includegraphics[width=0.8\linewidth]{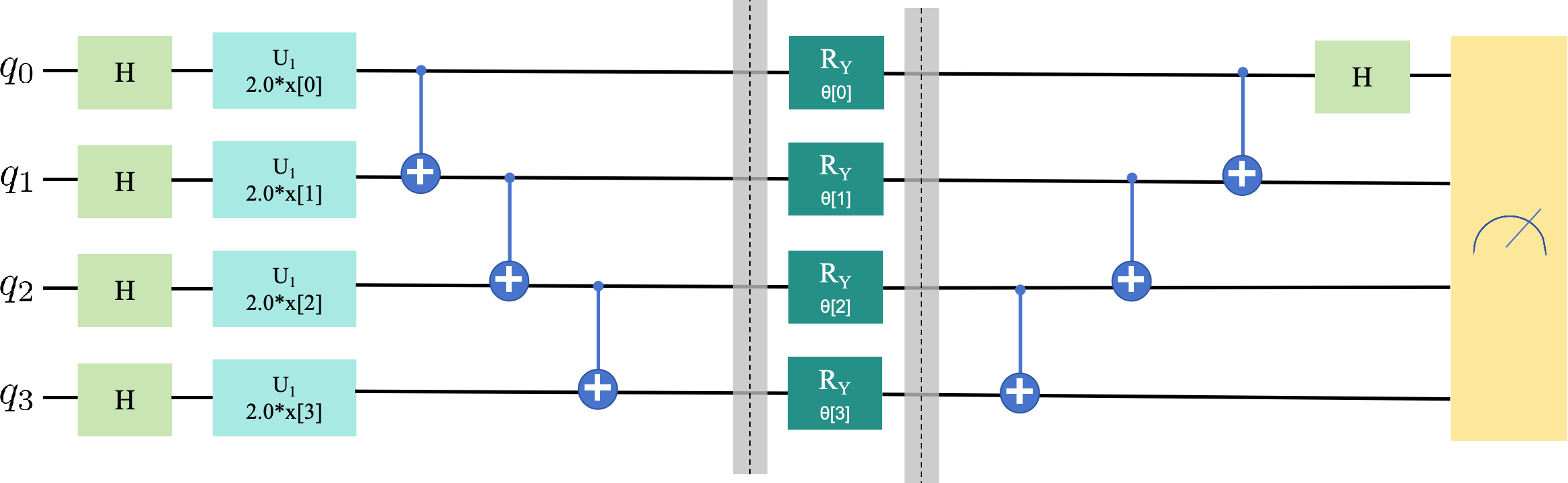}
     \caption{\label{fig:2}\textbf{Variational Quantum Circuit Structure.} The encoding layer of this network introduces H gates and U1 gates to each qubit, enabling adjustment and manipulation of individual qubit states by selecting appropriate rotation angles for the U1 gates. The research model integrates CNOT gates, a crucial multi-qubit operator that facilitates non-local state transfer by controlling interactions between the control and target qubits. }
\end{figure}

We employ quantum encoding techniques to process classical data\cite{PhysRevA.110.052615}. Within the quantum deep learning model, the single-qubit unitary layer serves as the core structure for quantum state transformation, comprising a series of meticulously designed single-qubit operations\cite{PhysRevA.80.022319}. These operations include, but are not limited to rotation gates, phase gates, and Hadamard gates, which collectively act on qubits to facilitate efficient processing and learning of quantum data. In the encoding layer of LCQHNN, we introduce Hadamard (H) gates and U1 gates for each qubit\cite{PhysRevLett.125.180504}. By appropriately selecting the rotation angle of the U1 gate in conjunction with the superposition characteristics of the H gate, we can effectively adjust and manipulate individual qubit states. Additionally, our model incorporates the controlled-not (CNOT) gate-a key multi-qubit operator that enables non-local state transfer through interactions between control and target qubits\cite{PhysRevLett.107.117203}. A notable feature of the CNOT gate is its parameter-free operation it relies solely on the current states of both qubits when they are in a superposition state. The application of this CNOT gate results in entanglement between them, introducing powerful non-linear characteristics into quantum computation. Through this innovative encoding structure, as depicted in Figure \ref{fig:2}, the LCQHNN not only achieves efficient quantum encoding of classical data but also fully leverages principles from quantum mechanics such as superposition and entanglement, significantly enhancing both its learning capacity and expressive power.

After the data have undergone the encoding phase, they proceed to a crucial transition step, the parametrized layer, which is composed of tunable quantum gates that can perform delicate adjustments and transformations on the quantum state. The core function of the parametrized layer is to capture and enhance the intrinsic features of the input data through the evolution of quantum states. Within this layer, each qubit undergoes a rotation around the Y-axis, achieved by applying RY gates. The parameters of the RY gates are trainable, determining the proximity of the qubits to the \(|0\rangle\) and \(|1\rangle\) basis states in the quantum state space\cite{PhysRevA.103.052425}. The modulation of this proximity directly affects the model's representational power, significantly influencing the depth and breadth of feature extraction. These parameters can be precisely adjusted through classical optimization algorithms to suit specific classification tasks. During the training process, the algorithm iteratively optimizes the parameters to minimize the discrepancy between the model's predictions and the actual labels. This process not only enhances the model's sensitivity to data features but also strengthens its capability to address complex classification problems. The VQCs algorithmic procedure is formulated
in Equations (2-7).

We first apply four Hadamard gates (H) to transform the four qubits from their initial state $| 0 \rangle $ to a uniformly stacked state $ | \psi_1 \rangle $. Among them, Hadamard transforms $| 0 \rangle $ into $ \frac {1} {\sqrt {2}} (| 0 \rangle+| 1 \rangle) $\cite{PhysRevLett.61.483}.
\begin{equation}
     |\psi_1\rangle = H^{\otimes 4} |0\rangle^{\otimes 4}
\end{equation}

Then four single qubit rotation gates $U_{1, i} (\delta i) $ are applied to each qubit of $| \psi_1 \rangle $. The rotation angle $\delta _i $ is twice the input data $x [i] $. This operation rotates the phase of each quantum bit by $\delta_i $ radians, thereby generating a new quantum state $| \psi_2 \rangle $.
\begin{equation}
     |\psi_2\rangle = U_{1,0}(\delta _0) \otimes U_{1,1}(\delta _1) \otimes U_{1,2}(\delta _2) \otimes U_{1,3}(\delta _3) |\psi_1\rangle  (\delta_i = 2 \times x[i],i=1,2,3,4)
\end{equation}

Here  three controlled NOT gates (CNOT gates) are designed in VQCs for the application of $| \psi_2 \rangle $. The CNOT gate $CX_{i, j} $ is controlled by the i-th quantum bit and targets the j-th quantum bit. This operation entangles the information between quantum bits together, resulting in a new quantum state $| \psi_3 \rangle $.
\begin{equation}
      |\psi_3\rangle = CX_{0,1} \cdot CX_{1,2} \cdot CX_{2,3} |\psi_2\rangle
\end{equation}

Afterwards, we apply four single qubit Y-axis rotation gates $RY (\eta_i) $ to each qubit of $| \psi_3 \rangle $. The rotation angle $\eta_i $ is a model parameter. This operation causes the state of each quantum bit to rotate by an angle of $\theta_i $ on the Y-axis of the Bloch sphere, thereby changing the phase and amplitude of the quantum state.
\begin{equation}
      |\psi_4\rangle = RY(\theta_i)^{\otimes 4}|\psi_3\rangle (i=1,2,3,4)
\end{equation}

Then, we apply a series of CNOT gates to the $| \psi_4 \rangle $ state, in order from right to left. This means that the rightmost quantum bit (3rd quantum bit) serves as the control bit first, followed by the 2nd quantum bit, and finally the 1st quantum bit. This sequential CNOT gate operation can be seen as an effect of "canceling" previous CNOT gate operations, as they are applied in the opposite order.
\begin{equation}
      |\psi_5\rangle = CX_{2,3} \cdot CX_{1,2} \cdot CX_{0,1} |\psi_4\rangle
\end{equation}

Finally, we apply Hadamard gates $ H_0 $ to the first qubit of the $| \psi_5 \rangle $ state, while keeping the other qubits unchanged (using unit gates $I$). This operation is typically used in the final stage of quantum circuits to prepare for measurements.
\begin{equation}
      |\psi_6\rangle = H_0 \otimes I \otimes I \otimes I |\psi_5\rangle
\end{equation}

\begin{figure}[ht!]
\centering
\includegraphics[width=0.6\linewidth]{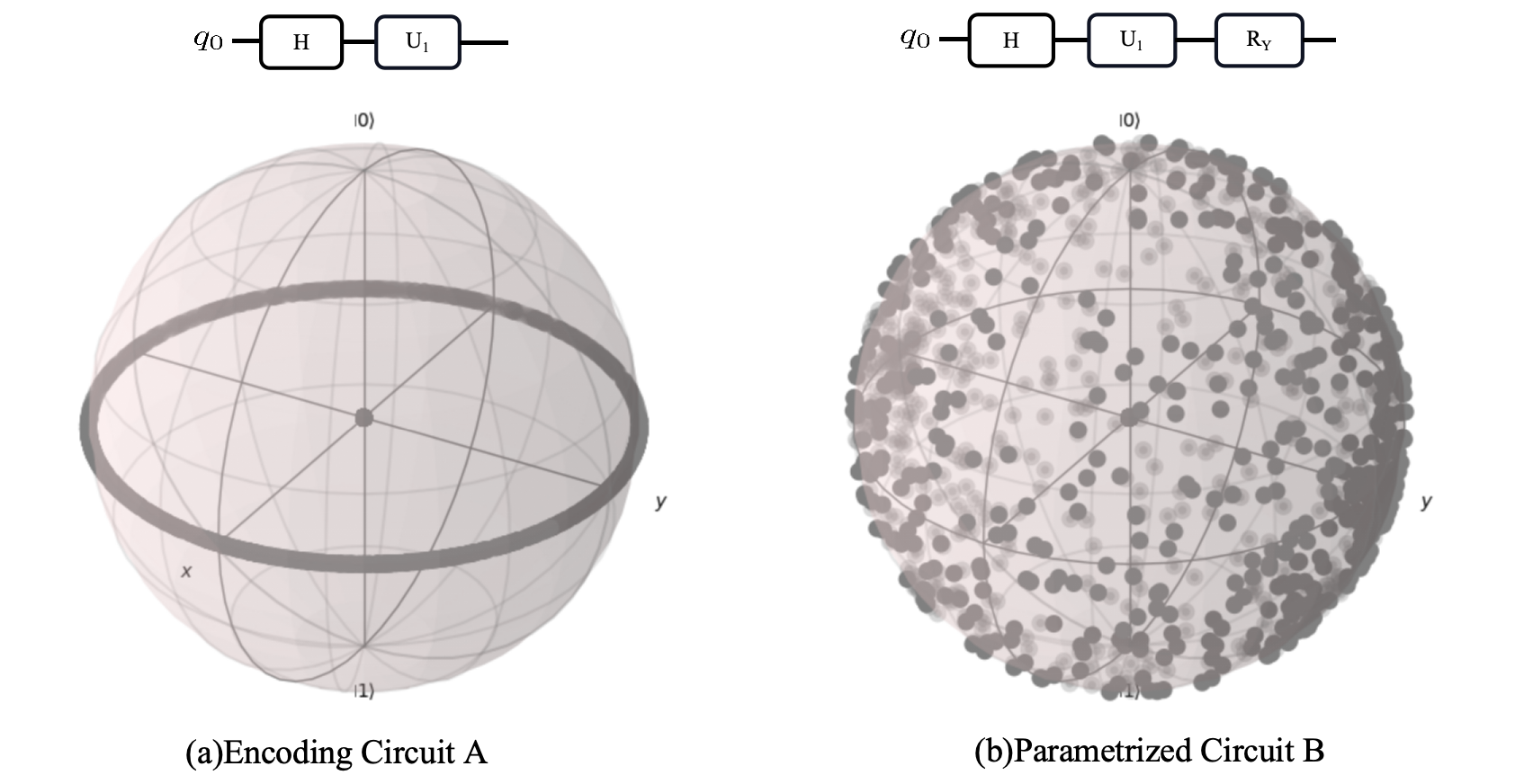}
\caption{\label{fig:3}\textbf{Bloch sphere representation of encoded quantum states.} (a) After the input qubit undergoes the encoding circuit with H and U1 gates, it becomes uniformly distributed on the equatorial plane of the Bloch sphere, at this point, the qubit is in a superposition state of \(|0\rangle\) and \(|1\rangle\). (b) The image displays the distribution of the quantum state after rotation around the y-axis. The parameter of the RY gate determines the final position of the quantum state on the Bloch sphere. }
\end{figure}

To observe the impact of quantum gates and operations on the quantum state more intuitively, we have mapped the output states of the two circuits onto the Bloch sphere. The specific distribution of quantum states is shown in Figure \ref{fig:3}. After the input bits pass through the H gates and U1 gates of the encoding circuit, they are uniformly distributed on the equatorial plane. At this point, the quantum bits are in an equal superposition state of the 0 state and 1 state. The uniform distribution of quantum states implies that the quantum information is uniformly encoded in the quantum state space, which is beneficial for quantum entanglement and quantum error correction. Moreover, the uniform distribution of quantum states also indicates that our encoding strategy does not introduce any specific preferences when processing sample data. This is crucial for enhancing the model's generalization capability. Such uniformity helps ensure that the model does not overfit to specific features in the training data, thereby maintaining high predictive accuracy and robustness when facing new, unseen data. Figure \ref{fig:3}(b) illustrates the distribution of quantum states after rotation around the Y-axis. The parameters of the RY gate determine the final position of the quantum states on the Bloch sphere. Since the Bloch sphere can only represent a single quantum bit, a uniform distribution does not mean that there is no specific entanglement formed between the quantum bits.

\section{Result}
\subsection{Preparations}
\subsubsection{Experimental Environment}
We conducted experiments on a server with a Xeon Gold 5315Y CPU and RTX3090 GPU, using Python 3.8.1, Pytorch 2.1.2, CUDA 11.8, Qiskit 1.0.2\cite{Paszke2019,qiskit2024}. Qiskit is an open-source quantum computing software development kit (SDK) developed and maintained by the IBM Quantum team. It allows users to design, simulate, validate, and run quantum programs and algorithms on their simulators or real quantum computers. Qiskit provides a complete set of tools for constructing quantum circuits, analyzing quantum information, and implementing quantum algorithms.

Our experiment uses Qiskit to simulate quantum circuits on a classical computer and integrates Qiskit with Pytorch through the Aer plugin. This integration enables the direct use of PyTorch's tensor operations in quantum circuits, as well as the utilization of PyTorch's automatic differentiation function in quantum algorithms.

\subsubsection{Datasets}
In order to evaluate the performance of the proposed image classification model, the MNIST, Fashion MNIST, and CIFAR-10 datasets were selected for experiments.

MNIST (Modified National Institute of Standards and Technology database) is a large handwritten digit database created by Yann LeCun et al. in the 1990s \cite{LeCun1998GradientBasedLearning}. It contains 60,000 training samples and 10,000 test samples, each of which is a $28 \times 28$ pixel handwritten digit image ranging from 0 to 9. The images in the MNIST dataset are grayscale and have been centered and standardized to ensure that the format of each image is consistent. Due to the large amount of computing resources and time required for training and validation processes, we have chosen a subset of MNIST, which includes 2048 balanced training samples and 512 balanced validation samples, respectively ("0" and "1"). We selected 1024 random samples as the test dataset to ensure the broad applicability of our method.

FashionMNIST is a dataset provided by Zalando (a German fashion e-commerce platform) as an alternative to MNIST \cite{xiao2017fashion}, aiming to provide a similar dataset structure but containing clothing images rather than handwritten digits. FashionMNIST contains 70,000 images, divided into 10 categories (such as T-shirts, pants, shoes, etc.), with 7000 images per category and a size of $28 \times 28$ pixels. In this study, we trained on 2048 images labeled with two categories ("pants" and "shirt"), validated with 512 images, and tested with 1024 images.

Additionally, we used the CIFAR-10 dataset \cite{2009Learning}, which consists of 60,000 $32 \times 32$ color images divided into 10 classes, with 6,000 images per class. For our binary classification task, we selected two classes ("airplane" and "car") from CIFAR-10. The training set included 2048 images, the validation set included 512 images, and the test set included 1024 images. This selection allowed us to further evaluate the model's performance on more complex, color images while maintaining a balanced dataset.

\subsubsection{Training Setting}
After converting the raw image data into feature maps, we trained the model using the Adam optimizer for 50 epochs with a batch size of 64 and a learning rate of 0.001. We determined a random seed of 42 to ensure the reproducibility of the algorithm.

\subsection{Classification Analysis of Multiple Datasets}

To substantiate and evaluate the effectiveness of the methodologies proposed within this manuscript, we have selected accuracy (Acc) as the evaluative criterion to measure the performance of disparate models. The computation of accuracy is defined by Equation (2), where N denotes the total number of classes, T signifies the total number of samples, the true class of each sample is represented by \(y_{i}\), and the class predicted by the model is indicated by \(\hat{y_{i}} \)\cite{Fisher1936THEUO}.

\begin{equation}
     Acc=\frac{ {\textstyle \sum_{i=1}^{N}}\Pi (\hat{y_{i}}=y_{i} ) }{T} \times 100\%
\end{equation}

At the same time, we also compared the convergence speed of different models. Assuming $E_{A}$ represents the number of epochs for model A to converge and $E_{B}$  represents the number of epochs for model B to converge, the percentage improvement $Q_{AB}$  in the convergence speed of model A compared to model B can be calculated using the following formula.

\begin{equation}
     Q_{AB} = \left( \frac{E_B - E_A}{E_B} \right) \times 100\%
\end{equation}

\begin{table}[ht!]
     \centering
     \caption{Model Architecture Comparison}
     \label{tab:0}
     \begin{tabular}{lcc}
     \toprule
     \textbf{Model} & \textbf{Architecture Description} & \textbf{Number of parameters} \\
     \midrule
     LCQHNN &
     \begin{tabular}[c]{@{}c@{}}Variational Quantum Circuit (VQC) \\ $\rightarrow$ FC(4$\rightarrow$2)\end{tabular} &
     $\begin{aligned} 4 + (4 \times 2) \\ = 8_{\text{weights}} \end{aligned}$ \\
     \midrule
     CNN-4 &
     \begin{tabular}[c]{@{}c@{}}FC(4$\rightarrow$1) \\ $\rightarrow$ FC(1$\rightarrow$2)\end{tabular} &
     $\begin{aligned} (4 \times 1) + (1 \times 2) \\ = 6_{\text{weights}} \end{aligned}$
     \\
     \midrule
     CNN-8 &
     \begin{tabular}[c]{@{}c@{}}FC(4$\rightarrow$2) \\ $\rightarrow$ FC(2$\rightarrow$2)\end{tabular} &
     $\begin{aligned} (4 \times 2) + (2 \times 2) \\ = 12_{\text{weights}}\end{aligned}$
      \\
     \midrule
     CNN-16 &
     \begin{tabular}[c]{@{}c@{}}FC(4$\rightarrow$4) \\ $\rightarrow$ FC(4$\rightarrow$2)\end{tabular} &
     $\begin{aligned} (4 \times 4) + (4 \times 2) \\ = 24_{\text{weights}}\end{aligned}$
      \\
     \bottomrule
     \end{tabular}
\end{table}

\begin{figure}[ht!]
\centering
\includegraphics[width=0.8\linewidth]{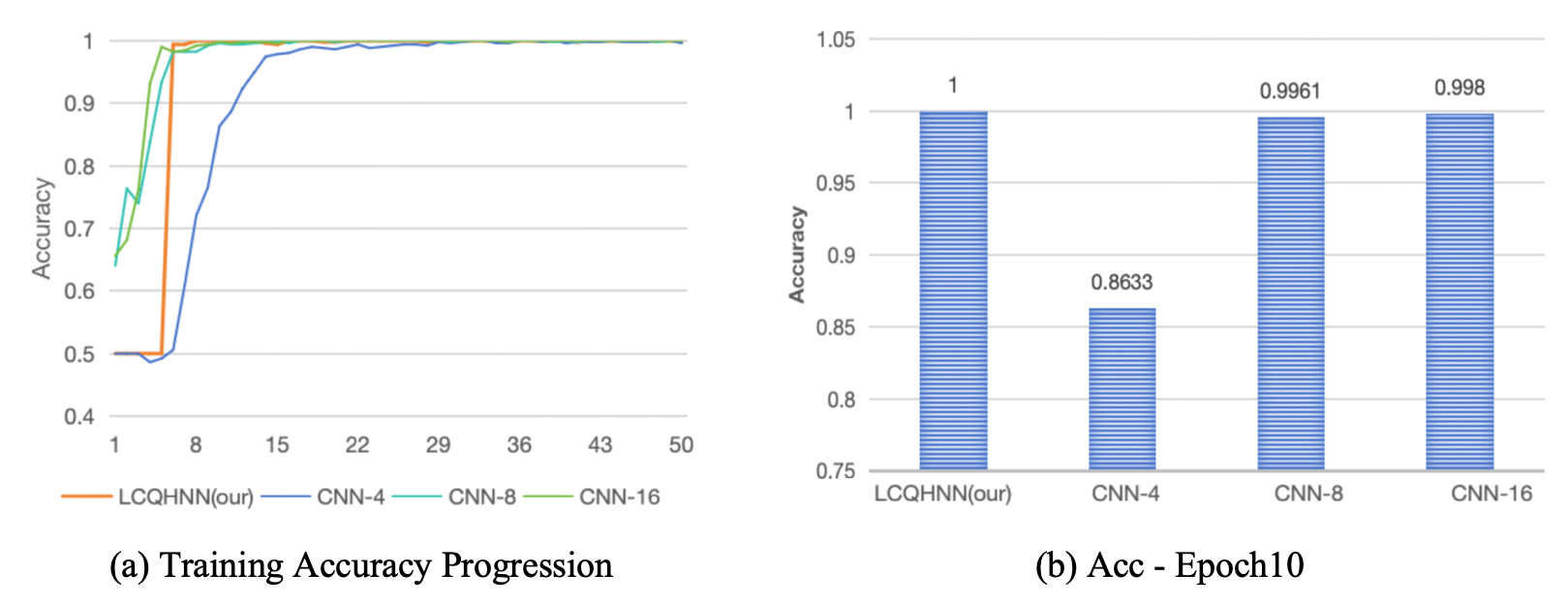}
\caption{\label{fig:4}\textbf{Experimental results on the MNIST dataset.} (a) The comparison of the LCQHNN against traditional CNNs with varying numbers of parameters in terms of accuracy evolution during the training process was conducted following the unified setup described in the paper. (b) The accuracy of different models at the 10th epoch during training. }
\end{figure}

We compared the LCQHNN against CNNs featuring fully connected modules with 4, 8, and 16 parameters, tracking the evolution of Accuracy throughout the training process. The classification module differences between models are shown in the Table \ref{tab:0}, where the numerical suffix (4/8/16) denotes the parameter count of the first FC layer in the classical models. As can be seen from Figure \ref{fig:4}(a), LCQHNN converges faster than a traditional cnn with 4 variable parameters, and the accuracy exceeds 99\% by the sixth epoch, which is 75\% quicker than its CNNs counterpart with the same parameters. Traditional CNNs necessitate a larger number of parameters to match this performance. Figure \ref{fig:4}(b) displays the accuracy of the four models after 10 epochs of training. It is apparent that the model introduced in this paper performs on par with, and occasionally exceeds, the capabilities of sophisticated traditional neural network models on the digit classification task. This hints at the VQCs's superior linear transformation capacity in comparison to the classical fully connected layers, allowing the model to concentrate on finer details, thereby attaining higher predictive precision. The data suggest that the LCQHNN possesses greater potential compared to conventional models.

\begin{figure}[ht!]
\centering
\includegraphics[width=0.8\linewidth]{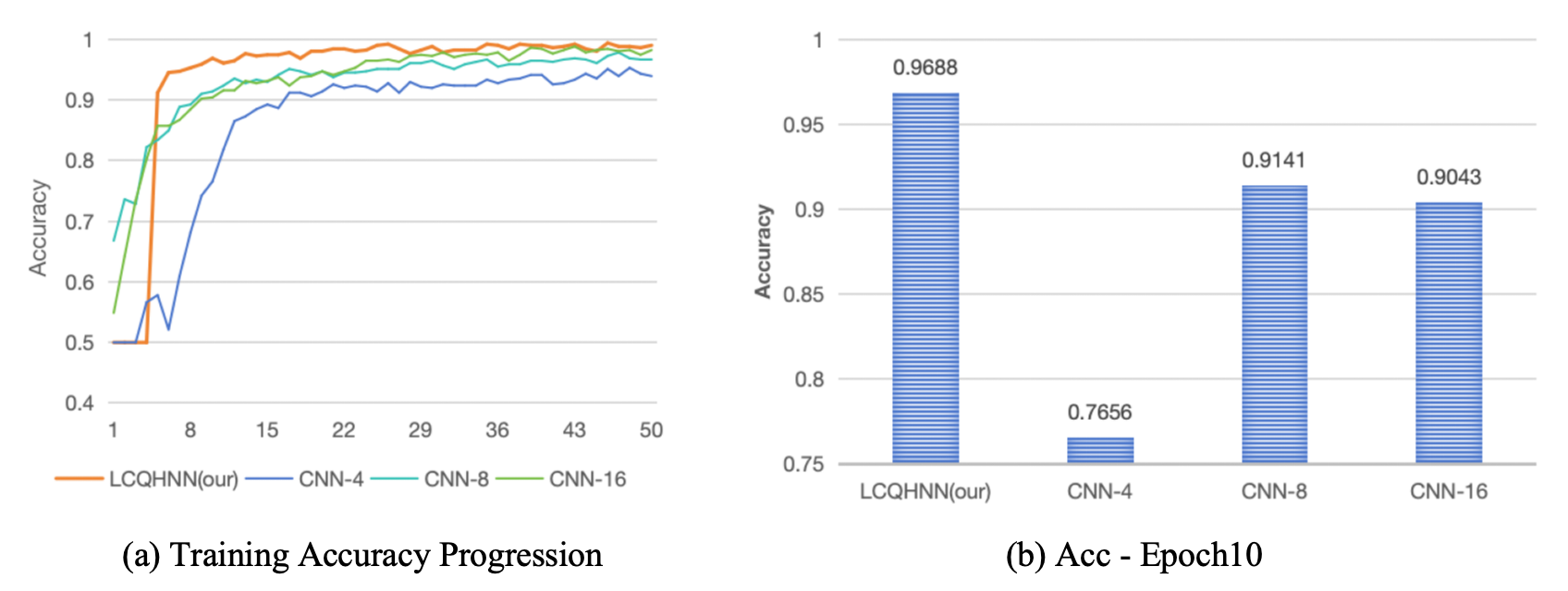}
\caption{\label{fig:5}\textbf{Experimental results on the FashionMNIST dataset.} (a) Comparison of the accuracy change of LCQHNN training on the FashionMNIST dataset with traditional CNNs with different numbers of parameters. (b) The accuracy of each model on the FashionMNIST dataset after training for 10 epochs. }
\end{figure}

Based on the previous experimental results, we trained four models on the more challenging FashionMNIST dataset and evaluated their classification accuracy on a separate test set, analogous to the structure of the previously discussed MNIST dataset. The experimental results are presented in Figure \ref{fig:5}. Figure \ref{fig:5}(a) illustrates the fluctuation in accuracy throughout the training process for all four models, with LCQHNN's rapid convergence being particularly evident. The LCQHNN exceeded an accuracy threshold of 90\% on the test set after only five epochs of training, representing a 70.59\% improvement in convergence speed compared to traditional CNN models with an equivalent number of parameters. Even models with a larger array of parameters, such as CNN-8 and CNN-16, did not demonstrate such notable efficacy. This phenomenon can be attributed to VQCs's adeptness at exploring high-dimensional feature spaces, facilitating quicker comprehension and adaptation to the underlying characteristics of the data. Figure \ref{fig:5}(b) delineates each models precision on the FashionMNIST test set after ten epochs these outcomes substantiate that within an identical constrained training period, LCQHNN achieves superior classification precision, corroborating that incorporating quantum attributes significantly enhances classification proficiency within a reduced training duration.

\begin{table}
     \centering
     \caption{\label{tab:1}Comparison of classification accuracy on public datasets.}
     \resizebox{\textwidth}{!}{
     \begin{tabular}{ccccccc}
     \toprule
     Model & MNIST (10) & MNIST (25) & MNIST (50) & FashionMNIST (10) & FashionMNIST (25) & FashionMNIST (50) \\

     \midrule
     LCQHNN (our) & \textbf{100\%} & 100\% & 100\% & \textbf{96.88\%} & \textbf{98.24\%} & \textbf{99.02\%} \\
     CNN-4 & 86.33\% & 99.22\% & 100\% & 76.56\% & 91.41\% & 93.95\% \\
     CNN-8 & 99.61\% & 100\% & 100\% & 91.41\% & 95.12\% & 96.68\% \\
     CNN-16 & 99.8\% & 100\% & 100\% & 90.43\% & 96.48\% & 98.24\% \\
     \bottomrule
     \end{tabular}
     }
\end{table}

The complete training data is presented in Table \ref{tab:1}. The MNIST dataset is relatively straightforward and is primarily used to explore the quantum characteristics within the LCQHNN. In terms of experimental evaluation metrics, there is no significant difference compared with traditional CNNs. However, on the more complex FashionMNIST dataset, it is evident that the LCQHNN outperforms the other three traditional CNNs of varying complexities in both convergence speed and final accuracy. The results indicate that the final test accuracy of the LCQHNN on FashionMNIST is 0.78\% higher than that of CNN-16, 2.34\% higher than CNN-8, and shows a significant improvement of 5.07\% over CNN-4.

To further validate the generalizability of the LCQHNN model, we systematically compared it with current mainstream image classification architectures in benchmark tests. Among them, HQNN (Hybrid Quantum Neural Network)\cite{qiskit2024}, as a similar hybrid architecture, is collaboratively constructed with classical convolutional modules and variational quantum circuits (VQCs), and its design paradigm originates from the cutting-edge research of the Qiskit machine learning team\footnote{Code available at: \url{https://qiskit-community.github.io/qiskit-machine-learning/tutorials/05_torch_connector.html}}.
QNN (Quantum Neural Network)\cite{farhi2018classificationquantumneuralnetworks}, on the other hand, is a fully quantum circuit-based end-to-end architecture that does not integrate any classical computational units\footnote{Code available at: \url{https://quict-docs.readthedocs.io/aa/latest/tutorials/algorithm/QNN}}.
By introducing the CIFAR-10 datasets, which contains complex textural features, this experiment aims to verify whether LCQHNN can maintain its architectural advantages when faced with visual tasks exhibiting significant cross-domain feature variations. As shown in Table \ref{tab:2}, the experimental results demonstrate that in the binary classification task on CIFAR-10, our proposed LCQHNN achieves the best performance (85.55\%) compared to CNN-16, HQNN, and QNN.
Except for QNN, all other models achieve perfect accuracy on MNIST, highlighting the necessity of classical-quantum hybridization. On FashionMNIST, although HQNN achieves slightly higher accuracy than LCQHNN (99.12\% vs. 99.02\%), this result does not overshadow LCQHNN's innovative architecture and its exceptional performance on CIFAR-10. The minor difference on FashionMNIST may be attributed to the dataset's lower complexity, which aligns well with HQNN's design. In contrast, LCQHNN's ability to excel in both simple and highly complex tasks underscores its potential as a versatile and future-proof solution for advanced image classification challenges.

\begin{table}
     \centering
     \caption{\label{tab:2}Performance comparison of LCQHNN with classical, quantum, and hybrid models on benchmark datasets.}
     \resizebox{0.6\textwidth}{!}{
     \begin{tabular}{cccc}
     \toprule
     Model & MNIST & FashionMNIST  & CIFAR-10 \\
     \midrule
     LCQHNN (our) & 100\% & 99.02\% & \textbf{85.55\%} \\
     CNN-16 & 100\% & 98.24\% & 83.89\% \\
     HQNN & 100\% & \textbf{99.12}\% & 83.79\% \\
     QNN & 85.93\% & 78.13\% & 60.93\%  \\
     \bottomrule
     \end{tabular}
     }
\end{table}

\begin{figure}[ht!]
     \centering
     \includegraphics[width=0.8\linewidth]{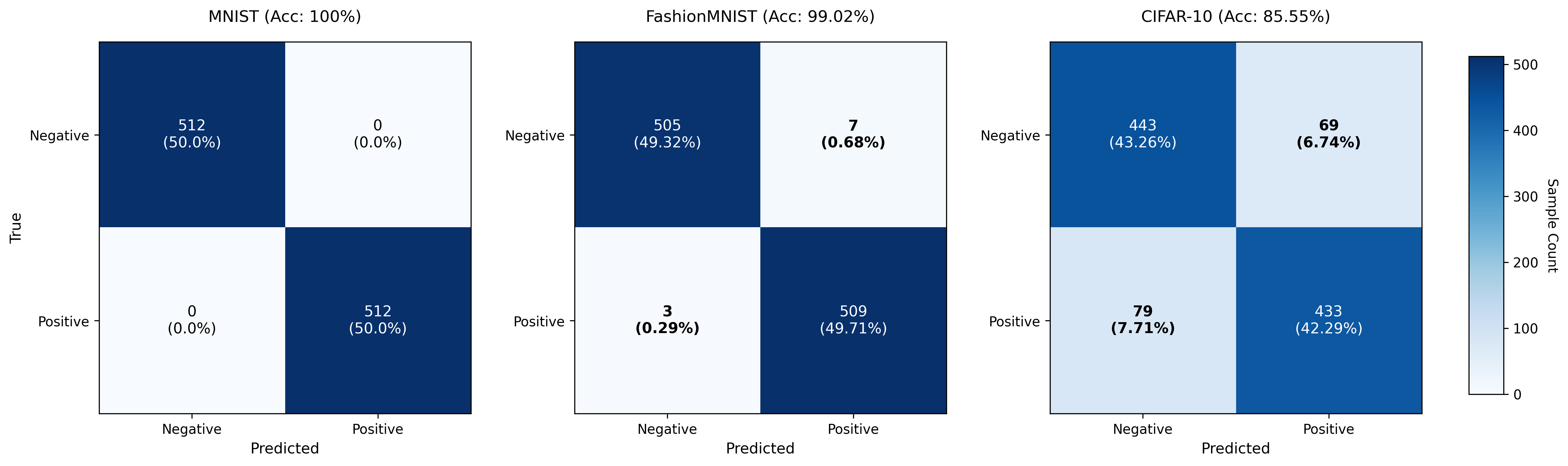}
     \caption{\label{fig:8}\textbf{The confusion matrices of LCQHNN on two distinct datasets.} In the binary classification task of MNIST (left), the model achieved 100\% accuracy. On the Fashion-MNIST dataset (middle) and CIFAR-10 dataset, minor classification errors were observed. }
\end{figure}

Figure \ref{fig:8} illustrates the confusion matrices of LCQHNN across three datasets, highlighting its classification performance on varying image complexities. In the MNIST binary task, the model achieved 100\% accuracy with zero misclassifications: all 512 negative samples ("0") and 512 positive samples ("1") were perfectly categorized. For Fashion-MNIST, the model attained 99.02\% accuracy, with minor errors including 7 false positives (pants misclassified as shirts) and 3 false negatives (shirts misclassified as pants). On the more challenging CIFAR-10 subset, LCQHNN achieved 85.55\% accuracy in distinguishing airplanes from cars. While correctly identifying 443 airplanes (TN) and 433 cars (TP), it exhibited 69 false positives (airplanes misclassified as cars) and 79 false negatives (cars misclassified as airplanes), reflecting the dataset's inherent complexity in fine-grained feature discrimination. These results demonstrate LCQHNN's robustness in simple classification tasks and its competitive capability in complex scenarios, with errors primarily arising from subtle inter-class variations in advanced vision tasks.

Although LCQHNN demonstrates promising performance on the Fashion-MNIST and CIFAR-10 datasets, its practical application is subject to several critical limitations. First, the experiments were conducted only on small-scale datasets ($\leq$70k samples), and the model's performance on larger-scale tasks (e.g., ImageNet) remains untested, potentially requiring architectural adaptations to handle increased data complexity. Second, the current evaluations rely on classical quantum simulators, which cannot fully replicate the noise characteristics of real quantum hardware. This discrepancy may lead to deviations between simulated results and actual performance in real-world quantum environments. Finally, it should be pointed out that our current model design is only for binary classification tasks, which to some extent limits its applicability to complex multi classification scenarios. We will investigate this issue in future work.

\subsection{Visual Explanations}

Neural network models are frequently considered as black-box systems due to their inherently complex nonlinear transformations and opaque internal structural parameters. This lack of transparency prompts researchers to investigate the focal points of these models. To enhance our understanding of the model's focus in this paper, we introduce Gradient-weighted Class Activation Mapping (Grad-CAM), an interpretability technique that leverages gradient information to identify regions in an image that contribute most significantly to the model's predictions for a specific class. By combining gradients with activation maps from convolutional layers to generate Class Activation Maps (CAMs), areas with substantial activation are superimposed onto the original image, thereby highlighting regions that play a critical role in class prediction. We visualized the training process of MNIST and FashionMNIST through heat maps, which provide clearer insights into shifts in the model's attention direction. We did not experiment on CIFAR-10 because the heat map visualization technique we used is specifically designed for grayscale images with a consistent pixel format. CIFAR-10 consists of color images with a different spatial resolution ($32 \times 32$ pixels) and channel structure (RGB), which makes it incompatible with our current visualization framework. The detailed computational formula is presented subsequently\cite{selvaraju2017gradcam}.

We first calculated the partial derivative of the model output $y$ with respect to the k-th convolutional layer output $A^{k}$. $y$ is usually the score of the model for a specific category. This partial derivative represents the contribution of each convolutional layer output to the final classification result.
\begin{equation}
      \alpha _{k}=\frac{\partial _{y}}{ \partial A^{k} }
\end{equation}

Then, on the feature map $A^{k}$ of the k-th convolutional layer, calculate the average of all partial derivatives $\alpha _{k}$ within the window centered on $(i, j)$. This average value $\omega _{k}(i, j)$ represents the average contribution of the k-th convolutional layer to the final classification result within the window.

\begin{equation}
     \omega _{k}(i, j) = \frac{1}{\text{Window Size}} \sum_{(m, n) \in \text{Window}(i, j)} \alpha _{k}(m, n)
\end{equation}

Finally, multiply the contribution degree $\omega _{k}$ of all convolutional layers by their corresponding feature maps $A^{k}$, and sum up all layers. Then, apply the ReLU function to the results to ensure that only positive contributions are retained. The final $G$ is a heatmap that displays the region in the input image that contributes the most to the classification results of a specific category.

\begin{equation}
     G=\max(0, \left (\textstyle \sum_{k}\omega _{k}A^{k}  \right ))
\end{equation}

\begin{figure}[ht!]
\centering
\includegraphics[width=0.45\linewidth]{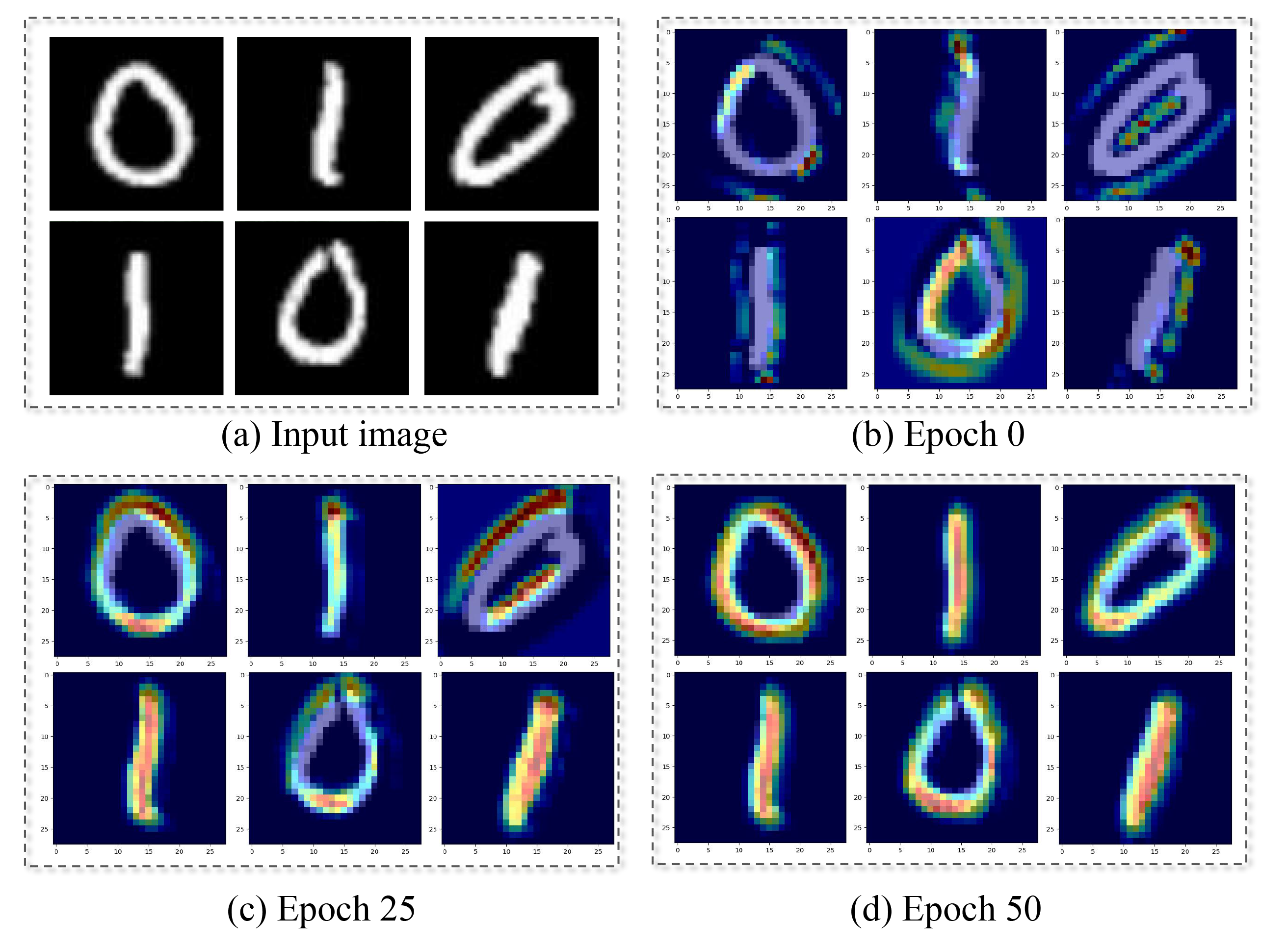}
\caption{\label{fig:6}\textbf{Visualization heatmap for model trained on MNIST dataset.} (a) Original Images Input to the Model. (b) Visualization of Untrained Model. (c) Visualization of Model Trained for 25 Epochs. (d) Visualization of Model Trained for 50 Epochs. }
\end{figure}

The visualization interpretation results of the LCQHNN on the MNIST dataset are depicted in \ref{fig:6}. We selected the last convolutional layer as the target layer to observe the significant areas the model uses for predictive concepts. Figure \ref{fig:6}(a) presents the input original image, while Figures 6(b), 6(c), and 6(d) illustrate the visualization interpretation results of the model at different training stages. We use the heatmaps to represent the importance of various regions in the image for the target category. The closer the heatmap pixel color is to dark red, the more critical the model considers that area for predicting the target category. Initially, for the untrained model (Figure \ref{fig:6}(b)), the visualization shows that the model's focus is scattered and chaotic, with no discernible pattern. This indicates that the model's initial cognition of the target category is inaccurate, and it has not yet learned effective features related to prediction. Subsequently, the model in the middle of training (Figure \ref{fig:6}(c)) begins to learn some feature information from the training set data, the model starts to focus on the edge contours of the handwritten digits, signifying that it is gradually understanding some important features of the digit shapes through training. Finally, the model after completion of training (Figure \ref{fig:6}(d)) shows that the areas of focus largely coincide with the shape of the handwritten digits. This demonstrates that the model has learned the feature information of the data and associated these features with the target categories after training. The model's ability to accurately identify and locate the important areas of the handwritten digits confirms the effectiveness of the LCQHNN on this task. Moreover, this indicates the practicality of the VQCs in neural networks, enhancing the model's expressive power and performance by combining classical and quantum computing advantages. By leveraging the unique features of quantum computing, VQCs can offer advantages in handling complex data patterns and tasks, providing superior predictive and interpretive capabilities.

\begin{figure}[ht!]
\centering
\includegraphics[width=0.45\linewidth]{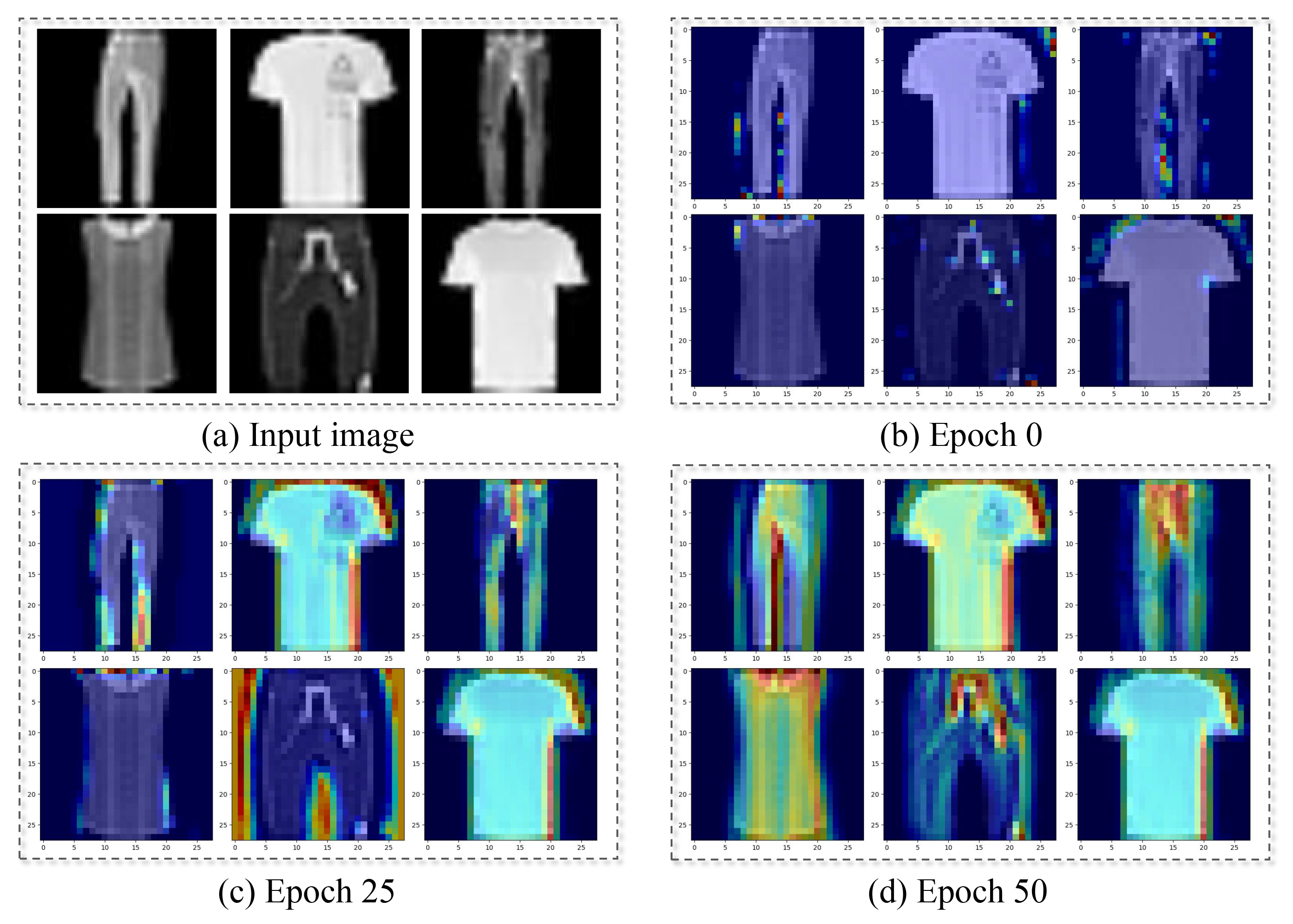}
\caption{\label{fig:7}\textbf{Visualization heatmap for model trained on FashionMNIST dataset.} (a) Original Images Input to the Model. (b) Visualization of Untrained Model. (c) Visualization of Model Trained for 25 Epochs. (d) Visualization of Model Trained for 50 Epochs. }
\end{figure}

Furthermore, our investigation extends to visualizing the decision-making process of the LCQHNN on the FashionMNIST dataset. As depicted in Figure \ref{fig:7}, the visualization outcomes allow us to distinctly perceive the progression of the model's focal points of attention. Initially, during the nascent training phase, the model's attention was divided, and as the training progressed, the model gradually redirected its attention to the key components of the image, eventually accurately identifying the object in question. Nonetheless, contrary to the MNIST dataset, the heatmaps on FashionMNIST do not manifest a saturated phenomenon even after successive training iterations, a discrepancy potentially owing to the visual intricacy inherent in FashionMNIST images. By juxtaposing the heatmaps that evolve with each training epoch against the accuracy curves, a pronounced correlation emerges. As the model's concentration on the predictive targets intensifies, there is a concomitant rise in test accuracy. This correlation substantiates that the model indeed seizes upon critical image data features throughout the training regimen, with these features being instrumental in the amelioration of classification proficiency. The visual analytics within this research not only offer a clear vantage point for deciphering the LCQHNN's decision-making apparatus but also underscore the model's capacity to refine its performance through attentive readjustments during the learning trajectory when confronted with sophisticated image data. These revelations furnish persuasive evidence for delving deeper into the nascent potential of quantum computation within the realm of deep learning, thereby establishing a groundwork for the innovation of hybrid quantum-classical models that are both more efficacious and precise.

\section{Conclusion}
In this study, we propose an innovative hybrid model that integrates classical and quantum computing, named LCQHNN, and conduct an in-depth comparative analysis of it. The novelty of the LCQHNN model lies in its combination of the powerful feature extraction capabilities of traditional CNNs with the classification potential of VQCs. Specifically, the model first utilizes CNNs to extract feature information from images and then effectively hands off these features to VQCs for precise classification tasks. Experimental results demonstrate that the improved classical-quantum hybrid model exhibits superior performance in classification tasks across multiple public datasets, with enhancements in both recognition accuracy and convergence speed compared to traditional CNNs models with the same parameters. This means that the LCQHNN model combining classical networks and quantum computing can achieve higher accuracy and provide a reliable, flexible, and scalable deep learning method for image classification tasks. Additionally, we observed the model's visual interpretation results, which show that the model effectively captured key data features during the training process and established associations with corresponding categories. This visual analysis highlights the capability of LCQHNN to enhance performance by fine-tuning its learning trajectory when dealing with complex image data. The LCQHNN framework holds significant promise for practical deployment in scenarios requiring high-precision and efficient image analysis. For instance, in medical diagnostics, it could enhance the detection of subtle anomalies in radiological images (e.g., early-stage tumors in MRI scans) by leveraging quantum-enhanced classification to reduce false negatives. In autonomous systems, such as self-driving vehicles, LCQHNN's compactness and accuracy could improve real-time object recognition under dynamic environmental conditions. Despite its advantages, practical adoption of LCQHNN faces several challenges. First, current quantum hardware limitations, such as qubit decoherence, gate fidelity, and limited qubit counts, constrain the scalability of VQCs for large-scale datasets. Second, The LCQHNN's reliance on transferring classical features (extracted by CNNs) to quantum circuits introduces a critical bottleneck: the dimensionality mismatch between high-dimensional classical features and limited qubit resources. Current quantum encoders struggle to efficiently map complex image features (e.g., 512-dimensional CNN outputs) into low-qubit quantum states without significant information loss. This trade-off between encoding fidelity and quantum resource overhead may degrade performance in real-world scenarios with high-resolution inputs. As quantum hardware advances, LCQHNN could emerge as a transitional framework that harmonizes classical and quantum paradigms, advancing AI solutions in domains where the interplay of precision and efficiency is critical.

\section*{Acknowledgments}
This work was supported by the National Natural Science Foundation of China under Grants No. 12374408  and No. 12475051;  the Natural Science Foundation of Hunan Province under grant No. 2023JJ30384;  the science and technology innovation Program of Hunan Province under grant No. 2024RC1050; and the innovative research group of Hunan Province under Grant No. 2024JJ1006.

\textbf{Data availability:}
The datasets used and/or analysed during the current study available from the corresponding author on reasonable request.

\textbf{Code availability:}
The code and implementation details for this study are publicly available at \url{https://github.com/LAAAAAAAAA/LCQHNN}.

\newcommand{\JournalTitle}{}
\bibliographystyle{MSP}
\bibliography{main}

\end{document}